\documentclass[a4paper,11pt]{article}
\usepackage{pos}
\usepackage{enumitem}
\usepackage{wrapfig}
\usepackage{lineno}
\usepackage{hyperref}

\title{Data production and Monte Carlo simulations for the Pierre Auger Observatory}
\author[a]{Paula Gina Isar} 
\affiliation[a]{Institute of Space Science-Subsidiary of INFLPR, Atomistilor 409, 077125 Magurele, Romania}

\onbehalf{for the Pierre Auger Collaboration$^{b,**}$}
\affiliation[b]{Observatorio Pierre Auger, Av. San Martín Norte 304, 5613 Malargüe, Argentina}

\affiliation[**]{Full author list: \rm\url{https://www.auger.org/archive/authors_icrc_2025.html}}

\emailAdd{spokespersons@auger.org }
\abstract{The Pierre Auger Observatory consists of 1660 water-Cherenkov detectors (WCDs) and 27 fluorescence telescopes, covering a surface of 3000 km$^{2}$ in the province of Mendoza, Argentina. After almost two decades, Auger Phase~I has ended the data taking and has already delivered many outstanding physics results. The Auger Observatory is currently running in Phase II, with upgraded detectors for enhanced data of ultra-high-energy cosmic rays (UHECRs). During Phase~I, several engineering detectors were developed and tested. Some of those developments led to new detectors, which are now part of the upgraded Observatory. For Phase~II, the WCDs were upgraded with a surface-scintillator detector, a radio detector, a small photomultiplier, and an upgraded electronics board, facilitating mass-sensitive measurements for source identification of UHECRs. Moreover, an underground muon detector allows for a direct measurement of the muon content of air showers. The Offline framework is the main software for the event reconstruction and detector simulation at the Pierre Auger Observatory. It is continuously being developed to accommodate changes in the detector and new algorithms. It allows the running of various applications, which are made up of individual modules, either on raw data or simulations on an event-by-event basis. The modules can be selected, configured, and run without re-compiling the whole software framework, giving the Auger user the possibility to define and further develop targeted analysis. Reconstructed events are stored in a compact ROOT file for physics analysis. Released versions of the Offline software are available for local installation on several platforms and distributed via the CERN-Virtual Machine File System (CVMFS) through a container. Large productions of shower and detector reference libraries are done with CORSIKA 7 and Offline on the European Grid Infrastructure (EGI) using the Virtual Organization Auger. Various air-shower geometries (up to 89 degrees in zenith) were considered, as well as several primary particles (from light to ultra-heavy nuclei, photons, and neutrinos), hadronic interaction models and detector configurations for Phase~I and Phase~II. In this contribution, we present the actual efforts and updates on software for data production and reference Monte Carlo simulations for Auger data analysis. }

\ConferenceLogo{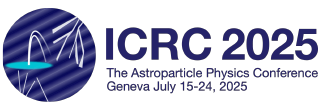}

\FullConference{39th International Cosmic Ray Conference (ICRC2025)\\
 15–24 July 2025\\
Geneva, Switzerland\\}
\begin{document}
\maketitle
%\linenumbers

\section{Introduction}
\begin{figure}[h]
    \centering
    \includegraphics[width=.524\linewidth]{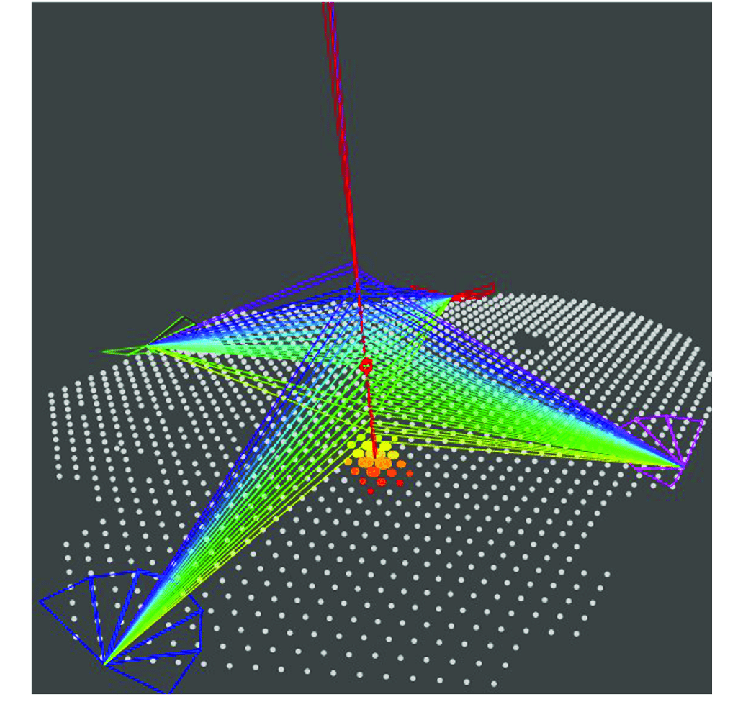}%
    \includegraphics[width=.42\linewidth]{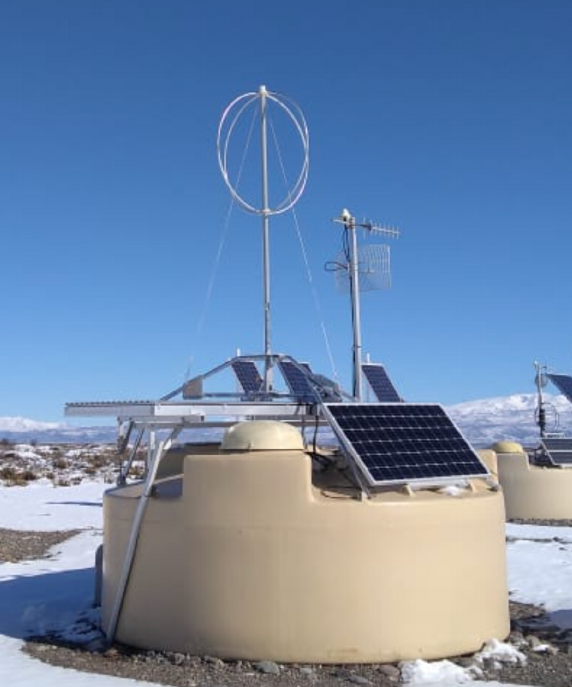}%
    \caption{A 3-D illustration of a hybrid event example measured at the Pierre Auger Observatory (left). The upgraded WCD of AugerPrime (right).}
    \label{fig:1}
\end{figure}

The Pierre Auger Observatory, located in the province of Mendoza, Argentina, is the world’s largest cosmic-ray detector. It measures the highest energetic particles in the Universe to elucidate one of the current open questions on their actual nature and origin. After over 20 years of data taking, Auger has ended Phase I and has already delivered many outstanding physics results, for instance \cite{APJ2024, EPJC2021}. The super-hybrid detector (Figure \ref{fig:1}, left) employs 1660 water-Cherenkov detectors (WCD) and 27 fluorescence detectors (FD) grouped in four locations, overlooking the entire array of 3000 km$^{2}$. Besides the main detection methods used to measure the ultra-high-energy cosmic rays (UHECRs), several atmospheric monitoring systems are also employed to ensure the quality of the acquired data. These consist of sensors for daily measurements of indicators such as temperature, pressure, humidity and wind \cite{JINST2012}, as well as lidars \cite{JINST2013} for measuring the cloud coverage and aerosol content based on a common measurement calendar together with the FDs, all serving in a full event reconstruction (where the response of the detector is corrected for various atmospheric conditions).

Currently, Auger is running in Phase II (AugerPrime), with upgraded detectors for enhanced data of UHECRs in increased statistics. The WCDs are upgraded (Figure~\ref{fig:1}, right) -  facilitating mass-sensitive measurements for source identification of UHECRs - with the following components:
\begin{itemize}
\item Surface-scintillator detector (SSD) \cite{POS392} and radio detector (RD) \cite{POS344, JCAP2018}, installed atop each WCD - improve the estimation of the mass of the primary particle. Both techniques are complementary since SSDs are sensitive to particle showers up to a zenith angle of 60 degrees, while for higher zenith angles, the RD starts to become efficient \cite{POS094};
\item Small photomultiplier tube (SPMT) \cite{POS343}, installed inside each WCD - serves to increase the dynamic range;
\item Upgraded unified board (UUB) \cite{POS230} electronics, replaced at each WCD - manages several devices, including front-end electronics, slow control, LED controller, GPS receiver, clock generator, memory and the various connectors from the SSD, RD and SPMT. It also accounts with a faster digitizer;
\item Underground muon detector (UMD), located at the sites of SD-750 and SD-433 (i.e., arrays with surface-detectors (SD) arranged in a triangular grid of either 750 m or 433 m spacing) – provides a direct measurement of the muon content of air showers. 
\end{itemize}

\section{Offline framework}

\begin{figure}
\begin{minipage}[b]{0.7\linewidth}
\centering
\includegraphics[scale=0.7]{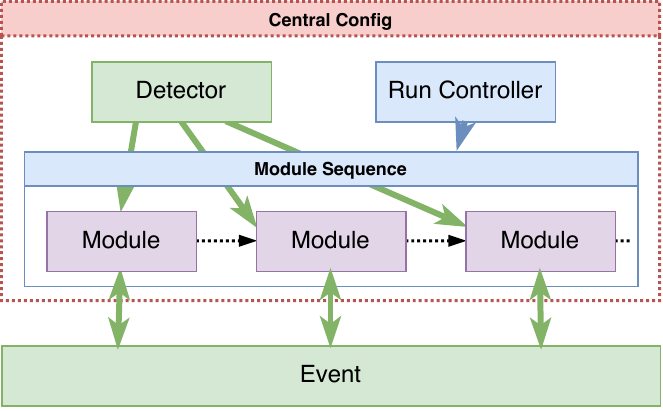}
\end{minipage}
\begin{minipage}[b]{0.8\linewidth}
\centering
\includegraphics[scale=0.62]{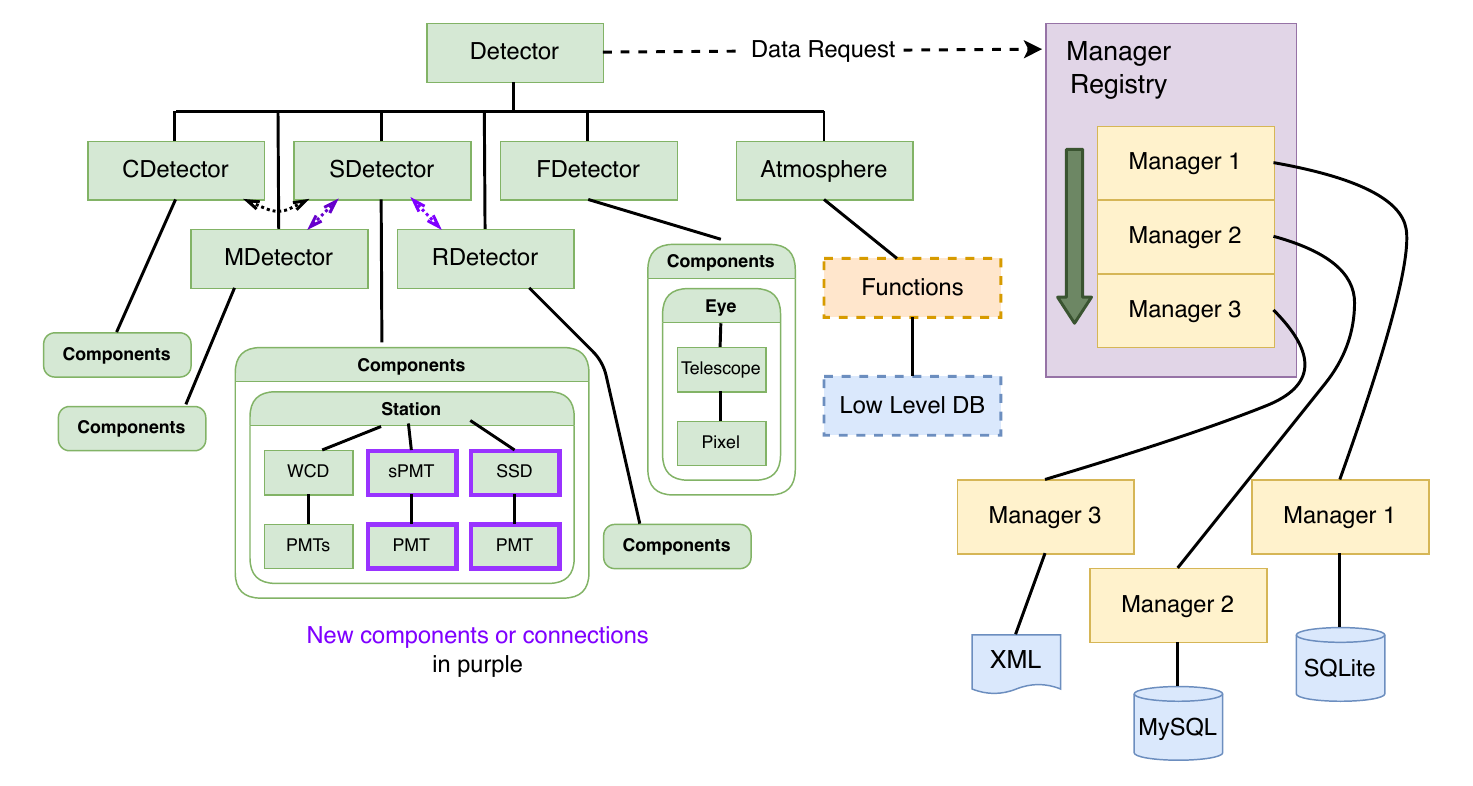}
\end{minipage}
\caption{The main structure of the Offline framework: Central Configuration and Event (top). The hierarchy of the Detector: front-end and the Manager interface: back-end (bottom). }
\label{fig:2}
\end{figure}

The Auger Offline framework \cite{NIMA2007, NIMA2011}, based on C++, is the main software used by the Pierre Auger Collaboration for the event reconstruction and simulation \cite{POS248}. Over 20 years, it has been continuously developed, yet to support the AugerPrime measurements with upgraded detectors \cite{POS353}. Moreover, it has been adopted by several other international projects. Its main structure, shown in the top panel of Figure \ref{fig:2}, comprises three components: 
\begin{itemize}
\item Detector---representing the detector components and time-varying atmospheric conditions;
\item Run Control---representing the module sequence;
\item Modules---accommodating the physics algorithms for processing, based on XML files, with static detector information. 
\end{itemize}

The Event, the principal communicator between the modules (Figure \ref{fig:2}, top panel), containing the central data structure \cite{POS353}, is divided into two main categories: the event type as measured (reconstructed and triggered data) and simulated data (energy, nuclear mass and direction of primary particle, longitudinal profile of developing air shower, etc.). Various applications based on individual modules have been developed at the detector level and are oriented to specific analysis to accommodate all the changes made over time with great flexibility. The structure of the detector and its main components are shown in the bottom panel of Figure \ref{fig:2}. These are the AugerPrime stations, consisting of the WCD, including the RD, SSD, and the SPMT, as well as the UMD modules, the FD, and the Auger Engineering Radio Array – AERA. Therefore, the Offline software framework supports the simulation and reconstruction of events of the surface (including radio measurements), fluorescence, and hybrid measurements, i.e., events measured by both surface and fluorescence detectors, as well as simulation of calibration techniques and other auxiliary applications. For the air-shower simulations, CONEX, CORSIKA 7/CoREAS, and AIRES/ZHAireS codes are considered for various types of detector arrays, such as SD-1500, SD-750, SD-433, or AERA \cite{POS113}, the proof-of-concept for the AugerPrime RD.
	
The Earth’s atmosphere is part of the detector and is therefore included as a component in its structure (Figure \ref{fig:2}, bottom-left). The time-varying atmospheric monitoring data and the calibration data are available in a low-level database (such as MySQL or SQLite), which serves the event reconstruction for atmospheric corrections and the response of the detectors under various atmospheric conditions. Atmospheric monitoring \cite{JINST2012, JINST2013} is performed at regular time intervals ranging from several minutes (for cloud coverage) or several hours (for aerosol monitoring) to several days (for vertical profiles of temperature, pressure, and humidity). Each detector component has a corresponding event structure, which gives the Auger developer/user the flexibility to design a dedicated module and/or a specific physics analysis. 
	
The Offline file format is either a ROOT \cite{NIMA1997} file for data analysis or a JSON file for Open Data \cite{EPJC2025} releases. Offline files contain the response of the several detectors employed in the event reconstruction, together with the main shower observables, such as the estimation of the cosmic-ray energy, depth of the shower maximum, and shower geometry. Typical Offline file sizes are in the order of a few tens of MB. 

\begin{figure}
\centering
\includegraphics[scale=0.6]{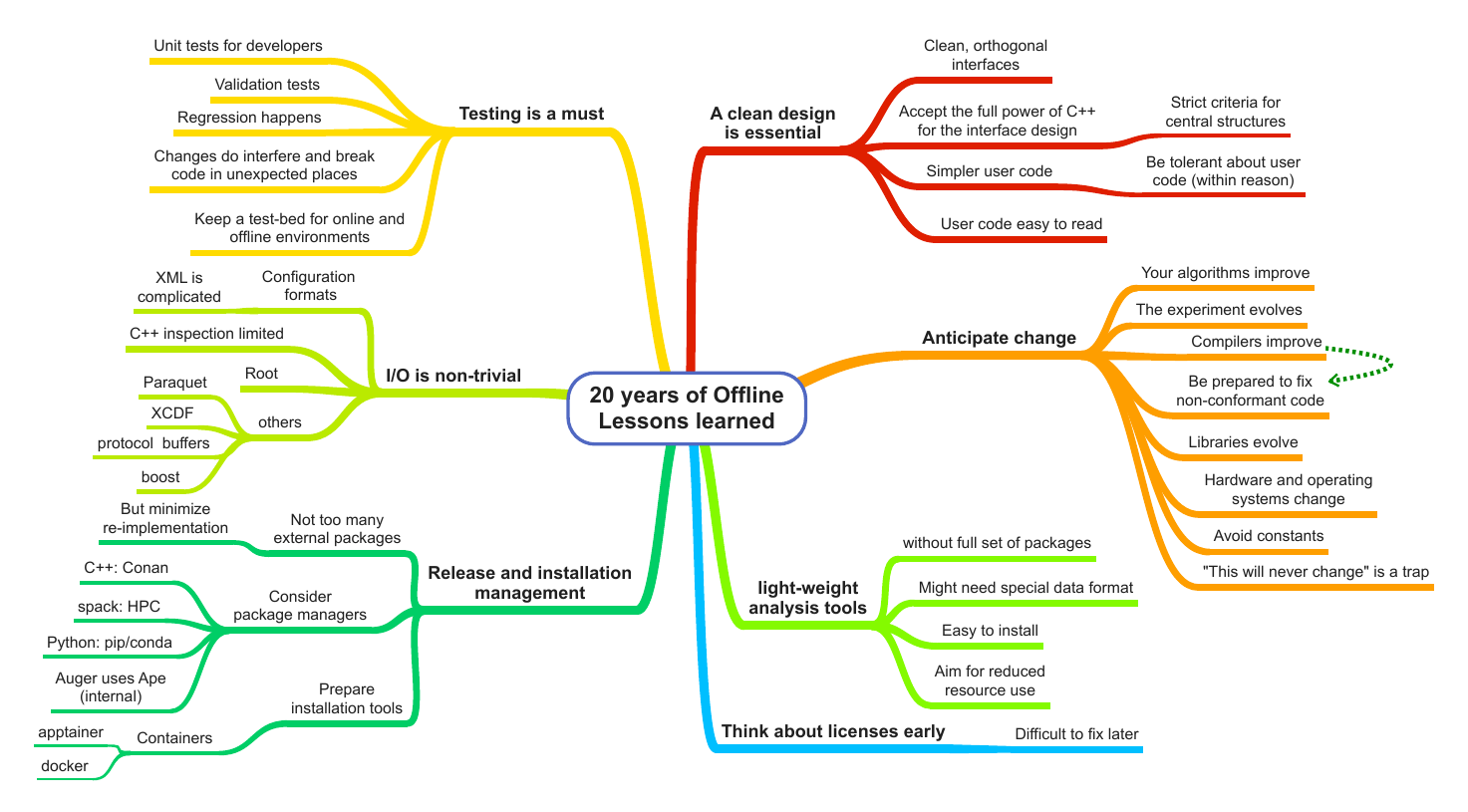}
\caption{Mind map with lessons learned on the Offline framework.
\label{fig:3}}
\end{figure}

So far, there are several lessons learned \cite{ARX2012}, which help both the developer and the user to, e.g., anticipate change, such as modifications to the detector (Figure \ref{fig:3}). 
	
Additionally, the evaluation of the open standard for machine learning (ML) operability, i.e., Open Neural Network Exchange (ONNX) is a work in progress for integrating ML into modern reconstruction and data analysis \cite{PRL2025, PRD2025} in a complementary and accelerated approach based on, e.g., deep-learning algorithms trained on large datasets of label data.

\section{Virtual Organization Auger}

The bulk production of extensive air-shower and detector reference libraries \cite{POS2021} is done in the grid via the Virtual Organization (VO) Auger. These libraries are standardized and were designed to serve as a base for a wide range of analyses, several of which resulted or will result in Collaboration publications. Also, specific analyses may require dedicated simulation libraries, which can also be produced by request. In the latter, priority is given to requests from the task coordinators, as well as the potential for publication. 

\begin{figure}
\centering
\includegraphics[scale=1.3]{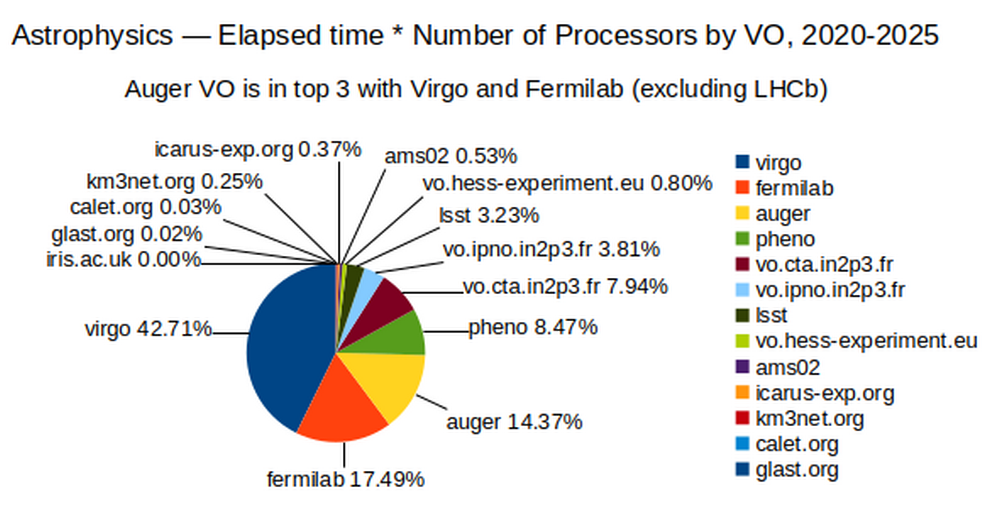}%VO-Auger-1706.png 0.65
\caption{The VOs in Astrophysics (without LHCb) showing the Elapsed-Time times Number of Processors by VO, in the last five years.
\label{fig:4}}
\end{figure}

VO Auger grid resources \cite{NIMA2007} have been used since 2006 \cite{JPCS2010} within an international effort of seven Auger member countries, contributing currently with ten grid sites for processing (and only a few grid sites in which we store the data) for distributed computing used in the mass production of library simulations (Figure \ref{fig:4}). The grid production is done via the European Grid Infrastructure using the DIRAC instance \cite{JPCS2010}, with job submission based on Python scripts. Preliminary tests, or pre-productions, are also done at the level of a local cluster (in the case of, e.g., CoREAS events with very large zenith angles, i.e., above 80 degrees, which usually may take as long as 30 days, if only one CPU is used). Released versions of Offline are available for local installation and distributed via the CERN-Virtual Machine File System (CVMFS) \cite{JPCS2015} through a container, which also serves in the production of reference libraries \cite{POS113}. Production files can either be accessed from the DIRAC File Catalog – DFC (for which an X509 certificate is required) or from the Auger Data Center at CNAF (central computing facility of INFN) – in progress. 

\section{Reference Monte Carlo simulations}

Most of our shower simulations use CORSIKA 7, from which a subset includes the radio content of air showers by activating the CoREAS option. The CORSIKA/CoREAS productions employ several hadronic interaction models combined with several atmospheric profiles for Malargüe (such as of January, March, August and September), and primary particle species at various primary energies, namely:
\begin{itemize}
\item p, He, O, Fe, including ultra-heavy elements (Te, Pt, U), and also photons. The energy range of these simulations is  $10^{14}$ $\leq$ E/eV $\leq$ $10^{20.5}$;
\item $\nu_{e}$(CC, NC), $\nu_{\tau}$(CC, Earth-skimming)-planned, E/eV > $10^{16}$;
\item QGSJETIII-01, EPOS LHC-R, SIBYLL 2.3e.
\end{itemize}

A typical CORSIKA 7 file size is large, in the order of 300 MB.

In a second step, the simulated showers are processed with the Auger Offline framework to generate several event libraries comprising the detection techniques and instrumentation available at the Auger site. These libraries consist of simulated events detected by the SD-1500, SD-750, and SD-433 arrays, as well as hybrid events detected both by the SD array and the FD, including the High Elevation Auger Telescopes (HEAT), for the low-energy extension of the Observatory. Lately, we have also provided AugerPrime simulations, consisting of multi-hybrid events containing the SSD and the UMD, together with the SD and the FD. In addition, information on the real-time state of the detectors and atmosphere is included in specific productions.

\section{Conclusions}

Auger Phase II is ongoing and will collect full-efficiency data for ten more years, aiming to provide high-statistics, mass-sensitive measurements to trace the origin of UHECRs. Remarkable physics results have been achieved so far with unprecedented measurement statistics of UHECRs. However, more achievements are yet to come with the upgraded detectors.

Throughout 20 years, the Offline framework underwent various extensions and upgrades, accommodating the upgraded detectors nowadays and paving the way for the integration of ONNX for modern ML algorithms applied to the complementary and timely reconstruction of physics data analyses. 

The reference Monte Carlo simulations serve for more efficient data analysis in the frame of the Pierre Auger Collaboration, which are targeted mainly for Collaboration papers and graduate academic studies. The CORSIKA 7 reference library production makes Auger the 3rd largest Astrophysics VO (with about 1 PB of data in DFC), reflecting the added value of distributed computing for accelerated processing of large volumes of data. 

So far, 30\% of measured data of the Pierre Auger Observatory will be released as Open Data \cite{POS646} for both the scientific community and the general public, thus contributing broadly to the awareness of scientific knowledge and STEM (science, technology, engineering and math) education in general, through a variety of resources available online.

\newpage
\section*{The Pierre Auger Collaboration}

{\footnotesize\setlength{\baselineskip}{10pt}
\noindent
\begin{wrapfigure}[11]{l}{0.12\linewidth}
\vspace{-4pt}
\includegraphics[width=0.98\linewidth]{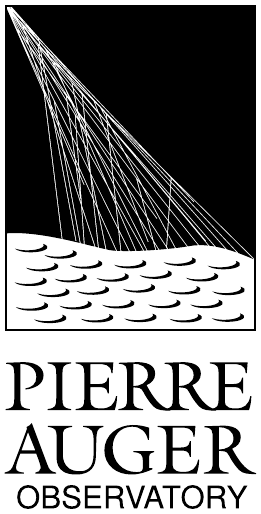}
\end{wrapfigure}
\begin{sloppypar}\noindent
% created on 2025-06-06
A.~Abdul Halim$^{13}$,
P.~Abreu$^{70}$,
M.~Aglietta$^{53,51}$,
I.~Allekotte$^{1}$,
K.~Almeida Cheminant$^{78,77}$,
A.~Almela$^{7,12}$,
R.~Aloisio$^{44,45}$,
J.~Alvarez-Mu\~niz$^{76}$,
A.~Ambrosone$^{44}$,
J.~Ammerman Yebra$^{76}$
% ...

\end{sloppypar}
\begin{center}
\end{center}

\vspace{1ex}
% created on 2025-06-06
% needs \usepackage{enumitem}
\begin{description}[labelsep=0.2em,align=right,labelwidth=0.7em,labelindent=0em,leftmargin=2em,noitemsep,before={\renewcommand\makelabel[1]{##1 }}]
\item[$^{1}$] Centro At\'omico Bariloche and Instituto Balseiro (CNEA-UNCuyo-CONICET), San Carlos de Bariloche, Argentina
\item[$^{2}$] Departamento de F\'\i{}sica and Departamento de Ciencias de la Atm\'osfera y los Oc\'eanos, FCEyN, Universidad de Buenos Aires and CONICET, Buenos Aires, Argentina
\item[$^{3}$] IFLP, Universidad Nacional de La Plata and CONICET, La Plata, Argentina
\item[$^{4}$] Instituto de Astronom\'\i{}a y F\'\i{}sica del Espacio (IAFE, CONICET-UBA), Buenos Aires, Argentina
\item[$^{5}$] Instituto de F\'\i{}sica de Rosario (IFIR) -- CONICET/U.N.R.\ and Facultad de Ciencias Bioqu\'\i{}micas y Farmac\'euticas U.N.R., Rosario, Argentina
\item[$^{6}$] Instituto de Tecnolog\'\i{}as en Detecci\'on y Astropart\'\i{}culas (CNEA, CONICET, UNSAM), and Universidad Tecnol\'ogica Nacional -- Facultad Regional Mendoza (CONICET/CNEA), Mendoza, Argentina
\item[$^{7}$] Instituto de Tecnolog\'\i{}as en Detecci\'on y Astropart\'\i{}culas (CNEA, CONICET, UNSAM), Buenos Aires, Argentina
\item[$^{8}$] International Center of Advanced Studies and Instituto de Ciencias F\'\i{}sicas, ECyT-UNSAM and CONICET, Campus Miguelete -- San Mart\'\i{}n, Buenos Aires, Argentina
\item[$^{9}$] Laboratorio Atm\'osfera -- Departamento de Investigaciones en L\'aseres y sus Aplicaciones -- UNIDEF (CITEDEF-CONICET), Argentina
\item[$^{10}$] Observatorio Pierre Auger, Malarg\"ue, Argentina
\item[$^{11}$] Observatorio Pierre Auger and Comisi\'on Nacional de Energ\'\i{}a At\'omica, Malarg\"ue, Argentina
\item[$^{12}$] Universidad Tecnol\'ogica Nacional -- Facultad Regional Buenos Aires, Buenos Aires, Argentina
\item[$^{13}$] University of Adelaide, Adelaide, S.A., Australia
\item[$^{14}$] Universit\'e Libre de Bruxelles (ULB), Brussels, Belgium
\item[$^{15}$] Vrije Universiteit Brussels, Brussels, Belgium
\item[$^{16}$] Centro Brasileiro de Pesquisas Fisicas, Rio de Janeiro, RJ, Brazil
\item[$^{17}$] Centro Federal de Educa\c{c}\~ao Tecnol\'ogica Celso Suckow da Fonseca, Petropolis, Brazil
\item[$^{18}$] Instituto Federal de Educa\c{c}\~ao, Ci\^encia e Tecnologia do Rio de Janeiro (IFRJ), Brazil
\item[$^{19}$] Universidade de S\~ao Paulo, Escola de Engenharia de Lorena, Lorena, SP, Brazil
\item[$^{20}$] Universidade de S\~ao Paulo, Instituto de F\'\i{}sica de S\~ao Carlos, S\~ao Carlos, SP, Brazil
\item[$^{21}$] Universidade de S\~ao Paulo, Instituto de F\'\i{}sica, S\~ao Paulo, SP, Brazil
\item[$^{22}$] Universidade Estadual de Campinas (UNICAMP), IFGW, Campinas, SP, Brazil
\item[$^{23}$] Universidade Estadual de Feira de Santana, Feira de Santana, Brazil
\item[$^{24}$] Universidade Federal de Campina Grande, Centro de Ciencias e Tecnologia, Campina Grande, Brazil
\item[$^{25}$] Universidade Federal do ABC, Santo Andr\'e, SP, Brazil
\item[$^{26}$] Universidade Federal do Paran\'a, Setor Palotina, Palotina, Brazil
\item[$^{27}$] Universidade Federal do Rio de Janeiro, Instituto de F\'\i{}sica, Rio de Janeiro, RJ, Brazil
\item[$^{28}$] Universidad de Medell\'\i{}n, Medell\'\i{}n, Colombia
\item[$^{29}$] Universidad Industrial de Santander, Bucaramanga, Colombia
\item[$^{30}$] Charles University, Faculty of Mathematics and Physics, Institute of Particle and Nuclear Physics, Prague, Czech Republic
\item[$^{31}$] Institute of Physics of the Czech Academy of Sciences, Prague, Czech Republic
\item[$^{32}$] Palacky University, Olomouc, Czech Republic
\item[$^{33}$] CNRS/IN2P3, IJCLab, Universit\'e Paris-Saclay, Orsay, France
\item[$^{34}$] Laboratoire de Physique Nucl\'eaire et de Hautes Energies (LPNHE), Sorbonne Universit\'e, Universit\'e de Paris, CNRS-IN2P3, Paris, France
\item[$^{35}$] Univ.\ Grenoble Alpes, CNRS, Grenoble Institute of Engineering Univ.\ Grenoble Alpes, LPSC-IN2P3, 38000 Grenoble, France
\item[$^{36}$] Universit\'e Paris-Saclay, CNRS/IN2P3, IJCLab, Orsay, France
\item[$^{37}$] Bergische Universit\"at Wuppertal, Department of Physics, Wuppertal, Germany
\item[$^{38}$] Karlsruhe Institute of Technology (KIT), Institute for Experimental Particle Physics, Karlsruhe, Germany
\item[$^{39}$] Karlsruhe Institute of Technology (KIT), Institut f\"ur Prozessdatenverarbeitung und Elektronik, Karlsruhe, Germany
\item[$^{40}$] Karlsruhe Institute of Technology (KIT), Institute for Astroparticle Physics, Karlsruhe, Germany
\item[$^{41}$] RWTH Aachen University, III.\ Physikalisches Institut A, Aachen, Germany
\item[$^{42}$] Universit\"at Hamburg, II.\ Institut f\"ur Theoretische Physik, Hamburg, Germany
\item[$^{43}$] Universit\"at Siegen, Department Physik -- Experimentelle Teilchenphysik, Siegen, Germany
\item[$^{44}$] Gran Sasso Science Institute, L'Aquila, Italy
\item[$^{45}$] INFN Laboratori Nazionali del Gran Sasso, Assergi (L'Aquila), Italy
\item[$^{46}$] INFN, Sezione di Catania, Catania, Italy
\item[$^{47}$] INFN, Sezione di Lecce, Lecce, Italy
\item[$^{48}$] INFN, Sezione di Milano, Milano, Italy
\item[$^{49}$] INFN, Sezione di Napoli, Napoli, Italy
\item[$^{50}$] INFN, Sezione di Roma ``Tor Vergata'', Roma, Italy
\item[$^{51}$] INFN, Sezione di Torino, Torino, Italy
\item[$^{52}$] Istituto di Astrofisica Spaziale e Fisica Cosmica di Palermo (INAF), Palermo, Italy
\item[$^{53}$] Osservatorio Astrofisico di Torino (INAF), Torino, Italy
\item[$^{54}$] Politecnico di Milano, Dipartimento di Scienze e Tecnologie Aerospaziali , Milano, Italy
\item[$^{55}$] Universit\`a del Salento, Dipartimento di Matematica e Fisica ``E.\ De Giorgi'', Lecce, Italy
\item[$^{56}$] Universit\`a dell'Aquila, Dipartimento di Scienze Fisiche e Chimiche, L'Aquila, Italy
\item[$^{57}$] Universit\`a di Catania, Dipartimento di Fisica e Astronomia ``Ettore Majorana``, Catania, Italy
\item[$^{58}$] Universit\`a di Milano, Dipartimento di Fisica, Milano, Italy
\item[$^{59}$] Universit\`a di Napoli ``Federico II'', Dipartimento di Fisica ``Ettore Pancini'', Napoli, Italy
\item[$^{60}$] Universit\`a di Palermo, Dipartimento di Fisica e Chimica ''E.\ Segr\`e'', Palermo, Italy
\item[$^{61}$] Universit\`a di Roma ``Tor Vergata'', Dipartimento di Fisica, Roma, Italy
\item[$^{62}$] Universit\`a Torino, Dipartimento di Fisica, Torino, Italy
\item[$^{63}$] Benem\'erita Universidad Aut\'onoma de Puebla, Puebla, M\'exico
\item[$^{64}$] Unidad Profesional Interdisciplinaria en Ingenier\'\i{}a y Tecnolog\'\i{}as Avanzadas del Instituto Polit\'ecnico Nacional (UPIITA-IPN), M\'exico, D.F., M\'exico
\item[$^{65}$] Universidad Aut\'onoma de Chiapas, Tuxtla Guti\'errez, Chiapas, M\'exico
\item[$^{66}$] Universidad Michoacana de San Nicol\'as de Hidalgo, Morelia, Michoac\'an, M\'exico
\item[$^{67}$] Universidad Nacional Aut\'onoma de M\'exico, M\'exico, D.F., M\'exico
\item[$^{68}$] Institute of Nuclear Physics PAN, Krakow, Poland
\item[$^{69}$] University of \L{}\'od\'z, Faculty of High-Energy Astrophysics,\L{}\'od\'z, Poland
\item[$^{70}$] Laborat\'orio de Instrumenta\c{c}\~ao e F\'\i{}sica Experimental de Part\'\i{}culas -- LIP and Instituto Superior T\'ecnico -- IST, Universidade de Lisboa -- UL, Lisboa, Portugal
\item[$^{71}$] ``Horia Hulubei'' National Institute for Physics and Nuclear Engineering, Bucharest-Magurele, Romania
\item[$^{72}$] Institute of Space Science, Bucharest-Magurele, Romania
\item[$^{73}$] Center for Astrophysics and Cosmology (CAC), University of Nova Gorica, Nova Gorica, Slovenia
\item[$^{74}$] Experimental Particle Physics Department, J.\ Stefan Institute, Ljubljana, Slovenia
\item[$^{75}$] Universidad de Granada and C.A.F.P.E., Granada, Spain
\item[$^{76}$] Instituto Galego de F\'\i{}sica de Altas Enerx\'\i{}as (IGFAE), Universidade de Santiago de Compostela, Santiago de Compostela, Spain
\item[$^{77}$] IMAPP, Radboud University Nijmegen, Nijmegen, The Netherlands
\item[$^{78}$] Nationaal Instituut voor Kernfysica en Hoge Energie Fysica (NIKHEF), Science Park, Amsterdam, The Netherlands
\item[$^{79}$] Stichting Astronomisch Onderzoek in Nederland (ASTRON), Dwingeloo, The Netherlands
\item[$^{80}$] Universiteit van Amsterdam, Faculty of Science, Amsterdam, The Netherlands
\item[$^{81}$] Case Western Reserve University, Cleveland, OH, USA
\item[$^{82}$] Colorado School of Mines, Golden, CO, USA
\item[$^{83}$] Department of Physics and Astronomy, Lehman College, City University of New York, Bronx, NY, USA
\item[$^{84}$] Michigan Technological University, Houghton, MI, USA
\item[$^{85}$] New York University, New York, NY, USA
\item[$^{86}$] University of Chicago, Enrico Fermi Institute, Chicago, IL, USA
\item[$^{87}$] University of Delaware, Department of Physics and Astronomy, Bartol Research Institute, Newark, DE, USA
\item[] -----
\item[$^{a}$] Max-Planck-Institut f\"ur Radioastronomie, Bonn, Germany
\item[$^{b}$] also at Kapteyn Institute, University of Groningen, Groningen, The Netherlands
\item[$^{c}$] School of Physics and Astronomy, University of Leeds, Leeds, United Kingdom
\item[$^{d}$] Fermi National Accelerator Laboratory, Fermilab, Batavia, IL, USA
\item[$^{e}$] Pennsylvania State University, University Park, PA, USA
\item[$^{f}$] Colorado State University, Fort Collins, CO, USA
\item[$^{g}$] Louisiana State University, Baton Rouge, LA, USA
\item[$^{h}$] now at Graduate School of Science, Osaka Metropolitan University, Osaka, Japan
\item[$^{i}$] Institut universitaire de France (IUF), France
\item[$^{j}$] now at Technische Universit\"at Dortmund and Ruhr-Universit\"at Bochum, Dortmund and Bochum, Germany
\end{description}

% created on 2025-06-06
\section*{Acknowledgments}

\begin{sloppypar}
The successful installation, commissioning, and operation of the Pierre
Auger Observatory would not have been possible without the strong
commitment and effort from the technical and administrative staff in
Malarg\"ue. We are very grateful to the following agencies and
organizations for financial support:
\end{sloppypar}

\begin{sloppypar}
Argentina -- Comisi\'on Nacional de Energ\'\i{}a At\'omica; Agencia Nacional de
Promoci\'on Cient\'\i{}fica y Tecnol\'ogica (ANPCyT); Consejo Nacional de
Investigaciones Cient\'\i{}ficas y T\'ecnicas (CONICET); Gobierno de la
Provincia de Mendoza; Municipalidad de Malarg\"ue; NDM Holdings and Valle
Las Le\~nas; in gratitude for their continuing cooperation over land
access; Australia -- the Australian Research Council; Belgium -- Fonds
de la Recherche Scientifique (FNRS); Research Foundation Flanders (FWO),
Marie Curie Action of the European Union Grant No.~101107047; Brazil --
Conselho Nacional de Desenvolvimento Cient\'\i{}fico e Tecnol\'ogico (CNPq);
Financiadora de Estudos e Projetos (FINEP); Funda\c{c}\~ao de Amparo \`a
Pesquisa do Estado de Rio de Janeiro (FAPERJ); S\~ao Paulo Research
Foundation (FAPESP) Grants No.~2019/10151-2, No.~2010/07359-6 and
No.~1999/05404-3; Minist\'erio da Ci\^encia, Tecnologia, Inova\c{c}\~oes e
Comunica\c{c}\~oes (MCTIC); Czech Republic -- GACR 24-13049S, CAS LQ100102401,
MEYS LM2023032, CZ.02.1.01/0.0/0.0/16{\textunderscore}013/0001402,
CZ.02.1.01/0.0/0.0/18{\textunderscore}046/0016010 and
CZ.02.1.01/0.0/0.0/17{\textunderscore}049/0008422 and CZ.02.01.01/00/22{\textunderscore}008/0004632;
France -- Centre de Calcul IN2P3/CNRS; Centre National de la Recherche
Scientifique (CNRS); Conseil R\'egional Ile-de-France; D\'epartement
Physique Nucl\'eaire et Corpusculaire (PNC-IN2P3/CNRS); D\'epartement
Sciences de l'Univers (SDU-INSU/CNRS); Institut Lagrange de Paris (ILP)
Grant No.~LABEX ANR-10-LABX-63 within the Investissements d'Avenir
Programme Grant No.~ANR-11-IDEX-0004-02; Germany -- Bundesministerium
f\"ur Bildung und Forschung (BMBF); Deutsche Forschungsgemeinschaft (DFG);
Finanzministerium Baden-W\"urttemberg; Helmholtz Alliance for
Astroparticle Physics (HAP); Helmholtz-Gemeinschaft Deutscher
Forschungszentren (HGF); Ministerium f\"ur Kultur und Wissenschaft des
Landes Nordrhein-Westfalen; Ministerium f\"ur Wissenschaft, Forschung und
Kunst des Landes Baden-W\"urttemberg; Italy -- Istituto Nazionale di
Fisica Nucleare (INFN); Istituto Nazionale di Astrofisica (INAF);
Ministero dell'Universit\`a e della Ricerca (MUR); CETEMPS Center of
Excellence; Ministero degli Affari Esteri (MAE), ICSC Centro Nazionale
di Ricerca in High Performance Computing, Big Data and Quantum
Computing, funded by European Union NextGenerationEU, reference code
CN{\textunderscore}00000013; M\'exico -- Consejo Nacional de Ciencia y Tecnolog\'\i{}a
(CONACYT) No.~167733; Universidad Nacional Aut\'onoma de M\'exico (UNAM);
PAPIIT DGAPA-UNAM; The Netherlands -- Ministry of Education, Culture and
Science; Netherlands Organisation for Scientific Research (NWO); Dutch
national e-infrastructure with the support of SURF Cooperative; Poland
-- Ministry of Education and Science, grants No.~DIR/WK/2018/11 and
2022/WK/12; National Science Centre, grants No.~2016/22/M/ST9/00198,
2016/23/B/ST9/01635, 2020/39/B/ST9/01398, and 2022/45/B/ST9/02163;
Portugal -- Portuguese national funds and FEDER funds within Programa
Operacional Factores de Competitividade through Funda\c{c}\~ao para a Ci\^encia
e a Tecnologia (COMPETE); Romania -- Ministry of Research, Innovation
and Digitization, CNCS-UEFISCDI, contract no.~30N/2023 under Romanian
National Core Program LAPLAS VII, grant no.~PN 23 21 01 02 and project
number PN-III-P1-1.1-TE-2021-0924/TE57/2022, within PNCDI III; Slovenia
-- Slovenian Research Agency, grants P1-0031, P1-0385, I0-0033, N1-0111;
Spain -- Ministerio de Ciencia e Innovaci\'on/Agencia Estatal de
Investigaci\'on (PID2019-105544GB-I00, PID2022-140510NB-I00 and
RYC2019-027017-I), Xunta de Galicia (CIGUS Network of Research Centers,
Consolidaci\'on 2021 GRC GI-2033, ED431C-2021/22 and ED431F-2022/15),
Junta de Andaluc\'\i{}a (SOMM17/6104/UGR and P18-FR-4314), and the European
Union (Marie Sklodowska-Curie 101065027 and ERDF); USA -- Department of
Energy, Contracts No.~DE-AC02-07CH11359, No.~DE-FR02-04ER41300,
No.~DE-FG02-99ER41107 and No.~DE-SC0011689; National Science Foundation,
Grant No.~0450696, and NSF-2013199; The Grainger Foundation; Marie
Curie-IRSES/EPLANET; European Particle Physics Latin American Network;
and UNESCO.
\end{sloppypar}

}

\end{document}